\documentclass[11pt]{article}
\usepackage[colorlinks=true,citecolor=blue,linkcolor=blue]{hyperref}
\usepackage{multirow}
\usepackage[left=2cm,right=2cm,top=2cm,bottom=2cm]{geometry}
\usepackage{cite}
\usepackage{psfrag}
\usepackage{graphicx}
\usepackage{graphics}
\usepackage{epsfig}
\usepackage{amsmath,amsfonts,amssymb}
\usepackage{pstcol,pst-fill,pst-grad}
\usepackage{euscript}
\usepackage{pstricks}
\usepackage{wrapfig}
\usepackage{slashed}
\usepackage[toc,page]{appendix}
\usepackage{here}

\date{}
\begin{document}
	\title{\vspace{-3cm}
		\hfill\parbox{4cm}{\normalsize \emph{}}\\
		\vspace{1cm}
		{ Relativistic elastic scattering of an electron by a muon in the
field of a circularly polarized electromagnetic wave}}
	\vspace{2cm}
	
\author{Y. Mekaoui$^{1,2}$, M. Jakha$^{1}$, S. Mouslih$^{1,3}$, B. Manaut$^1$, R. Benbrik$^4$ and S. Taj$^{1,}$\thanks{Corresponding author, E-mail: s.taj@usms.ma}  \\
		{\it {\small$^1$ Sultan Moulay Slimane University, Polydisciplinary Faculty,}}\\
		{\it {\small Laboratoire de Recherche en Physique $\&$ Sciences pour l'Ingénieur (LRPSI),}}\\ {\it {\small Equipe de Physique Moderne et Appliquée (PhyMA), Beni Mellal, 23000, Morocco.}}\\		
		{\it {\small$^2$Sultan Moulay Slimane University,
		Faculty of Sciences and Techniques,		
		Beni Mellal, 23000, Morocco.}}		\\				
	{\it {\small$^3$Faculty of Sciences and Techniques,
		Laboratory of Materials Physics (LMP),
		Beni Mellal, 23000, Morocco.}}		\\				
	{\it {\small$^4$ Polydisciplinary Faculty, Laboratoire de Physique Fondamentale et Appliquée, Sidi Bouzid, B.P. 4162 Safi, Morocco.}}			
	}
	
\maketitle
\begin{abstract}
Within the framework of quantum electrodynamics, the scattering of an electron by a muon in the presence of a circularly polarized monochromatic laser field is investigated theoretically in the first Born approximation. The expressions for the amplitude and the differential cross section are derived analytically by adopting the Furry picture approach in which the calculations are carried out using exact relativistic Dirac-Volkov functions. We begin by studying the process taking into account the relativistic dressing of only the electron without muon. Then, in order to reveal the effect of the muon dressing, we fully consider the relativistic dressing of the electron and muon together in the initial and final states. As a result, the differential cross section is significantly reduced by the laser field. We find that the effect of laser-dressing of muon becomes noticeable at laser field strengths greater than or equal to $10^{9}~\text{V cm}^{-1}$ and therefore must be taken into account. The influence of the laser field strength and frequency on the differential cross section and multiphoton process is revealed. An insightful comparison with the laser-free results is also included.\\
keywords: Laser assisted process, Dirac-Volkov formalism, QED calculations, Analytical calculations.
\end{abstract}
\section{Introduction}
Strong-field physics is the general research area of laser-matter interaction. It aims to stimulate and control ultrafast processes and understand the mechanism underlying them \cite{reiss2010}. New physics is arising from the interaction of intense laser fields with atoms, molecules and particles. Presently and due to the availability of high-power lasers, worldwide efforts are devoted to investigate theoretically various basic quantum electrodynamics (QED) processes in the presence of a strong electromagnetic fields. Collisions of charged particles in the presence of lasers have received much attention in recent decades due to their broad applications and their contribution to the fundamental understanding of atomic structure. Particularly, electron scattering processes have been crucial to the development of science, both theoretically and experimentally. Its importance in atomic and molecular physics has long been recognized. In parallel with the development of the high-power femtosecond lasers, it has become feasible to test experimentally these scattering processes and to observe multiphoton processes \cite{test1,test2,test3}. An overview of preliminary studies on laser-assisted scattering processes has been presented in some books by Faisal \cite{faisal}, Mittleman \cite{mittleman} and Fedorov \cite{fedorov}. The first well-studied processes, both analytically and numerically, was laser-induced Compton scattering \cite{compton}, followed by Mott scattering of a laser-dressed electron by the Coulomb field of an atomic nucleus \cite{mott}. Roshchupkin \textit{et al} developed the theory of the electron scattering on a nucleus in the field of two plane electromagnetic waves \cite{Roshchupkin1,Roshchupkin2,Roshchupkin3,Roshchupkin4}. Lebed' \textit{et al}  studied Mott scattering process in the presence of one \cite{Roshchupkin5} or two \cite{Roshchupkin6} laser pulses. Then, with increasing laser intensities, other scattering processes were investigated. Hrour \textit{et al} \cite{hrour} reported proton scattering by Coulomb potential in a circularly polarized laser field considering the Coulomb effect distortion. Laser-assisted electron-electron scattering was analyzed in the presence of a linearly polarized powerful laser field in \cite{panek2004}. Electron-positron scattering in the field of a light wave was studied in the works \cite{bhabha1,bhabha2}. Elastic electron-proton scattering in the presence of a circularly \cite{dahiri} or linearly \cite{liu2014,wang2019} polarized laser field has been also reported. In addition to QED processes, many authors have studied some processes of the electroweak theory such as the decay of particles \cite{baouahi,jakha} and the Higgs-boson production \cite{ouhammou} in the presence of a circularly polarized laser field. In our turn, as a contribution to the enrichment of the scientific literature with such theoretical studies of scattering processes in the presence of laser, we focus in this work on the study of the electron-muon scattering process within the framework of QED in the presence of a circularly polarized laser field. Elastic electron-muon scattering is one of the most basic processes in particle physics. It played an important role in the discovery of the muon in 1936, while measuring the collisions of muons in cosmic radiation with atomic electrons by the American physicists Carl D. Anderson and Seth Neddermeyer \cite{muondiscovery}. The electron-muon elastic scattering cross section was measured in the 1960s using accelerator-produced muons \cite{exper1,exper2,exper3}. Experimentally, a great project, MUonE project, is devoted to
measure the differential cross section (DCS) of electron-muon elastic scattering as a function of the squared momentum transfer in order to determine the leading hadronic contribution to the muon magnetic moment $\textsl{g}$-2 \cite{muoneproject}. This scattering process has been of great interest to researchers and has also been well studied theoretically previously. The QED radiative corrections to the electron-muon scattering cross section have been calculated a long time ago in \cite{kaiser,ambrosio,eriksson} and recently at next-to-leading order (NLO) in \cite{nlo}. New phenomena in the scattering of an electron/positron by a muon in the presence of a linearly polarized laser field are investigated in the first Born approximation by Du \textit{et al} in \cite{du2018,du2018p}. The same process has been studied in the field of an elliptically polarized plane electromagnetic wave in the resonant \cite{muon1} and nonresonant \cite{muon2} cases. Nonresonant scattering of an electron by a muon in the pulsed light field was investigated in general relativistic \cite{muon3} and nonrelativistic \cite{muon4} cases. In this paper, we will add new and relevant insights regarding the electron-muon scattering in the field of a plane, monochromatic and circularly polarized electromagnetic wave. Through this, our principal contribution is that we have considered the laser-dressing of muon and discussed its effect on the DCS. The paper is organized as follows. Firstly, in section~\ref{sec:2},
we give a concrete theoretical derivation of the laser-assisted DCS without dressing of muon in the first order of perturbation theory using the Dirac-Volkov functions. In section~\ref{dressing}, we consider the laser-dressing of muon which introduces new modifications on the DCS. The main results obtained in all cases are presented and discussed in section~\ref{sec:res}. Our conclusions are summarized in the final section. The metric tensor $g^{\mu\nu}=\text{diag}(1,-1,-1,-1)$ and natural units $\hbar=c=1$, where $c$ is the speed of light in vacuum, are used throughout this work. For all $k$, the bold style $\textbf{k}$ is reserved for vectors.
\section{Laser-assisted differential cross section}\label{sec:2}
\begin{figure}[h]
\centering
\includegraphics[scale=0.5]{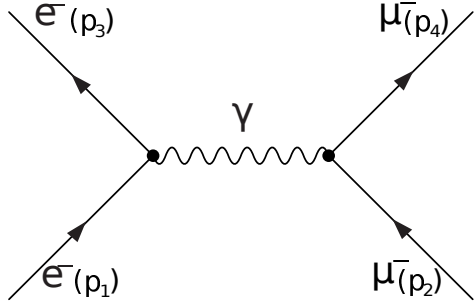}
\caption{Lowest Feynman diagram of electron-muon scattering in the framework of QED, where the intermediate propagator is a photon $\gamma$. Time flows from down to up.}\label{fig1}
\end{figure}

In this section, we will try to establish all the theoretical expressions of the quantities that are necessary to calculate the DCS of the electron-muon scattering process in the presence of a laser field. This process can be schematized as follows :
\begin{equation}\label{process}
e^{-}(p_{1})+\mu^{-}(p_{2})\longrightarrow e^{-}(p_{3})+\mu^{-}(p_{4}),
\end{equation}
where the labels $p_{i}$ are our associated 4-momenta. In the framework of QED, it can be described by the lowest Feynman diagram shown in figure~\ref{fig1}.
\\
We consider the applied laser as a plane, monochromatic and circularly polarized electromagnetic wave, theoretically defined by the following classical four-potential :
\begin{equation}\label{classical_fourpotential}
A^{\mu}(\phi)=a_1^\mu \cos(\phi)+a_2^\mu \sin(\phi),
\end{equation}
where $\phi=(k.x)$ is the phase of the laser field. $k=(\omega,\textbf{k})$ is the wave 4-vector and $\omega$ is the laser frequency. $a^{\mu}_{1}=(0,\textbf{a}_1)=|\text{\textbf{a}}|(0,1,0,0)$ and $a^{\mu}_{2}=(0,\textbf{a}_2)=|\text{\textbf{a}}|(0,0,1,0)$ are the polarization 4-vectors satisfying $(a_{1}.a_{2})=0$ and $a_{1}^{2}=a_{2}^{2}=a^{2}=-|\text{\textbf{a}}|^{2}=-(\mathcal{E}_{0}/\omega)^{2}$, where $\mathcal{E}_{0}$ is the electric field strength.  The classical four-potential $A^{\mu}(\phi)$ satisfies the Lorentz condition, $k_{\mu}A^{\mu}=0$, which implies $(k.a_{1})=(k.a_{2})=0$, meaning that the wave vector $\textbf{k}$ is chosen to be along the $z$-axis.\\
In the lowest order of perturbation theory, the scattering amplitude $S_{fi}$ for the laser-assisted electron-muon scattering can be written as \cite{greiner}
\begin{equation}\label{smatrix0}
S_{fi}=-ie^{2}\int d^{4}x\int d^{4}y~[\overline{\psi}_{e}^{f}(x)\gamma^{\mu}\psi_{e}^{i}(x)]D_{F}(x-y)[\overline{\psi}_{\mu^{-}}^{f}(y)\gamma_{\mu}\psi_{\mu^{-}}^{i}(y)],
\end{equation}
where $e=-|e|<0$ is the charge of the electron and $D_{F}(x-y)$ is the Feynman propagator for the electromagnetic radiation given by \cite{greiner}
\begin{equation}\label{Feynmanpropagator}
D_{F}(x-y)=\int\frac{d^{4}q}{(2\pi)^{4}}e^{-iq(x-y)}\bigg(\frac{-4\pi}{q^{2}+i\varepsilon}\bigg).
\end{equation}
At first, we will only consider the dressing of the electrons without the muons in the initial and final states. By adopting the Furry picture approach \cite{furry}, we take into account  the interaction of the initial and final electrons with the electromagnetic field by describing them by the relativistic Dirac-Volkov functions, which are the exact solutions of the Dirac equation in the presence of an electromagnetic field \cite{landau}. They are written, when normalized to the volume $V$, as follows \cite{volkov} :
\begin{equation}\label{ewaves}
\begin{split}
\psi_{e}^{i,f}(x)&=\bigg[1+\dfrac{e\slashed{k}\slashed{A}}{2(k.p_{1,3})}\bigg]\frac{u(p_{1,3},s_{1,3})}{\sqrt{2Q_{1,3}V}}\times e^{iS(q_{1,3},x)},
\end{split}
\end{equation}
where
\begin{equation}
\begin{split}
S(q_{1,3},x)&=-(p_{1,3}.x)-\int_{0}^{(k.x)}\bigg(\frac{e(p_{1,3}.A)}{(k.p_{1,3})}-\frac{e^{2}A^{2}}{2(k.p_{1,3})}\bigg)d\phi,\\
&=-(q_{1,3}.x)-\dfrac{e(a_1.p_{1,3})}{k.p_{1,3}}\sin(\phi)+\dfrac{e(a_2.p_{1,3})}{k.p_{1,3}}\cos(\phi).
\end{split}
\end{equation}
For the free initial and final muons, we have
\begin{equation}\label{muwaves}
\psi_{\mu^{-}}^{i,f}(y)=\frac{1}{\sqrt{2E_{2,4}V}}u(p_{2,4},s_{2,4})e^{-ip_{2,4}.y},
\end{equation}
where $u(p_{i},s_{i})$ represents the Dirac bispinor for the free electron and muon with momentum $p_{i}$ and spin $s_{i}$ satisfying $\sum_{s_{i}}u(p_{i},s_{i})\overline{u}(p_{i},s_{i})=\slashed{p}_{i}+m_{e/\mu}$, where $m_{e/\mu}$ is the rest mass of the electron or muon ($m_{\mu}\approx207m_{e}$). The laser-assisted kinetic 4-momentum of the electron is called effective momentum $q_{1,3}$, with the form
\begin{equation}
q_{1,3}=(Q_{1,3},\textbf{q}_{1,3})=p_{1,3}-\frac{e^{2}a^{2}}{2(k.p_{1,3})}k.
\end{equation}
Squaring this effective momentum shows that the mass of the dressed electron has a dependence on the field strength and frequency
\begin{equation}
q_{1,3}^{2}=m_{e*}^{2}=m_{e}^{2}-e^{2}a^{2},
\end{equation}
where $m_{e*}$ acts as an effective mass of the electron inside the laser field.\\
Substituting the expressions of Feynman propagator and wave functions into the scattering amplitude $S_{fi}$ (\ref{smatrix0}), we obtain
\begin{equation}\label{smatrix1}
\begin{split}
S_{fi}=&\dfrac{ie^2 4\pi}{\sqrt{16Q_{1}Q_{3}E_{2}E_{4}V^{4}}}\int d^4x d^4y \frac{d^4q}{(2 \pi)^4}\Big[ \frac{e^{i(q_3-q_1-q).x} e^{i(p_4-p_2+q).y}}{q^2+i\epsilon}\Big] ~e^{-iz \sin(\phi-\phi_0)}
\\
&\times \Big[\bar{u}(p_{3},s_{3})\Big(\chi_0^\mu+\chi_1^\mu\cos(\phi)+\chi_2^\mu\sin(\phi)\Big)u(p_{1},s_{1}) \Big]\Big[\bar{u}(p_{4},s_{4})\gamma_{\mu}u(p_{2},s_{2}) \Big],
\end{split}
\end{equation}
where
\begin{equation}\label{argument}
\begin{split}
z&=e\sqrt{\bigg(\dfrac{a_{1}.p_{1}}{k.p_{1}}-\dfrac{a_{1}.p_{3}}{k.p_{3}}\bigg)^{2}+\bigg(\dfrac{a_{2}.p_{1}}{k.p_{1}}-\dfrac{a_{2}.p_{3}}{k.p_{3}}\bigg)^{2}},\\
\phi_0&=\arctan\bigg[\frac{(a_{2}.p_{1})(k.p_{3})-(a_{2}.p_{3})(k.p_{1})}{(a_{1}.p_{1})(k.p_{3})-(a_{1}.p_{3})(k.p_{1})}\bigg],
\end{split}
\end{equation}
and
\begin{equation}
\begin{split}
&\chi_0^\mu=\gamma^{\mu}-e^{2}a^2k^\mu \slashed{k}/[2(k.p_{1})(k.p_{3})],\\
& \chi_1^\mu=[e/2(k.p_1)] \gamma^{\mu}\slashed{k}\slashed{a}_{1}+[e/2(k.p_3)]\slashed{a}_{1}\slashed{k}\gamma^{\mu},\\
& \chi_2^\mu=[e/2(k.p_1)] \gamma^{\mu}\slashed{k}\slashed{a}_{2}+[e/2(k.p_3)]\slashed{a}_{2}\slashed{k}\gamma^{\mu}.
\end{split}
\end{equation}
The integration over $d^4x$, $d^4y$ and $d^4q$ in eq.~(\ref{smatrix1}) can be performed by using standard techniques \cite{greiner}, and then we get
\begin{equation}\label{smatrix2}
\begin{split}
S_{fi}=& \dfrac{ie^2 4\pi}{\sqrt{16Q_{1}Q_{3}E_{2}E_{4}V^{4}}}
\sum_{s=-\infty}^{+\infty}
  \frac{(2\pi)^4 \delta^4 (p_4-p_2+q_3-q_1-sk)}{q^2+i \epsilon}
\\
&\times \Big[\bar{u}(p_{3},s_{3})\Big(\chi_0^\mu B_{s}(z)+\chi_1^\mu B_{1s}(z)+    \chi_2^\mu B_{2s}(z) \Big)u(p_{1},s_{1}) \Big]\Big[\bar{u}(p_{4},s_{4})\gamma_{\mu}u(p_{2},s_{2}) \Big],
\end{split}
\end{equation}
with $q=q_3 - q_1 -sk$ is the relativistic 4-momentum transfer in the presence of the laser field. In eq.~(\ref{smatrix2}), we have used the following transformation, known as Jacobi-Anger identity, involving ordinary Bessel functions \cite{landau}
\begin{align}\label{transformation}
\begin{bmatrix}
1\\
\cos(\phi)\\
\sin(\phi)
\end{bmatrix}\times e^{-iz\sin(\phi-\phi_{0})}=\sum_{n=-\infty}^{+\infty}\begin{bmatrix}
B_{s}(z)\\
B_{1s}(z)\\
B_{2s}(z)
\end{bmatrix}e^{-is\phi},
\end{align}
where $z$ is the argument of the Bessel function defined previously in eq.~(\ref{argument}) and $s$, its order, is interpreted as the number of photons exchanged between the electron and the laser field.\\
To express the laser-assisted DCS, we multiply the squared S-matrix element $|S_{fi}|^{2}$ by the density of final states, and divide by the observation time interval $T$ and the flux of the incoming particles $|J_{\text{inc.}}|$ and finally one has to average over the initial spins and to sum over the final ones. Then, we obtain
 \begin{equation}
 \begin{split}
 d\overline{\sigma}&= \frac{|S_{fi}|^2}{|J_{\text{inc.}}|T}\times \frac{Vd^3q_3}{(2\pi)^3}\times \frac{Vd^3p_4}{(2\pi)^3},\\
 &=\sum_{s=-\infty}^{+\infty} V^{2}\frac{d^{3}q_{3}}{(2\pi)^{3}}\frac{d^{3}p_{4}}{(2\pi)^{3}}\frac{e^{4}(4\pi)^{2}}{|J_{\text{inc.}}|T}\frac{(2\pi)^{4}VT\delta^{4}(p_4-p_2+q_3-q_1-sk)}{16Q_{1}Q_{3}E_{2}E_{4}V^{4}q^{4}}\frac{1}{4}\sum_{s_{i}}|\mathcal{M}_{fi}^s|^{2},\\
 &=\sum_{s=-\infty}^{+\infty} \frac{e^4\delta^4 (p_4-p_2+q_3-q_1-sk)}{16 Q_{1}Q_{3}E_{2}E_{4}|J_{\text{inc.}}| V q^{4}} d^3q_3 d^3p_4 \sum_{s_{i}}  |\mathcal{M}_{fi}^s|^2,
 \end{split}
 \end{equation}
where we have used the following property of the Dirac function \cite{greiner}:
\begin{align}
  [(2\pi)^{4}\delta^{4}(p_4-p_2+q_3-q_1-sk)]^{2}=(2\pi)^{4}VT\delta^{4}(p_4-p_2+q_3-q_1-sk),
  \end{align}
and
\begin{equation}
 \mathcal{M}_{fi}^s=\Big[\bar{u}(p_{3},s_{3})\Big(\chi_0^\mu B_{s}(z)+\chi_1^\mu B_{1s}(z)+    \chi_2^\mu B_{2s}(z) \Big)u(p_{1},s_{1}) \Big]\Big[\bar{u}(p_{4},s_{4})\gamma_{\mu}u(p_{2},s_{2}) \Big].
\end{equation}
With the help of the following formula \cite{greiner}:
\begin{equation}\label{formula0}
\frac{d^{3}p_{4}}{E_{4}}=2\int_{-\infty}^{+\infty}d^{4}p_{4}\delta(p_{4}^{2}-m_{\mu}^{2})\Theta(p_{4}^{0}),~~~\text{with}~~~\Theta(p_{4}^{0})=\left\lbrace\begin{array}{ccc}
&1&\text{for}~p_{4}^{0}>0\\
&0&\text{for}~p_{4}^{0}<0
\end{array}\right.
\end{equation}
the cross-section becomes
\begin{equation}
\begin{split}
d\overline{\sigma}&= \sum_{s=-\infty}^{+\infty} \frac{e^4}{8 E_{2}Q_{1}Q_{3}|J_{\text{inc.}}| V q^{4}} \int_{-\infty}^{+\infty} d^{4}p_{4} \delta^4 (p_4-p_2+q_3-q_1-sk)\delta(p_{4}^{2}-m_{\mu}^{2}) d^3q_3 \sum_{s_{i}}  |\mathcal{M}_{fi}^s|^2,\\
&=\sum_{s=-\infty}^{+\infty} \frac{e^4}{8 E_{2}Q_{1}Q_{3}|J_{\text{inc.}}| V q^{4}} \int\delta(p_{4}^{2}-m_{\mu}^{2}) d^3q_3 \sum_{s_{i}}  |\mathcal{M}_{fi}^s|^2\bigg|_{p_4-p_2+q_3-q_1-sk=0.}
\end{split}
\end{equation}
By considering the incident flux of electron in the rest frame of muon $|J_{\text{inc.}}|=\sqrt{(q_{1}.p_{2})^{2}-m_{\mu}^{2}m_{e*}^{2}}/(Q_{1} E_{2} V )=|\mathbf{q_1}|/(Q_1 V)$ and using $d^{3}q_{3}=|\textbf{q}_{3}|^{2}d|\textbf{q}_{3}|d\Omega_{f}$, we get
 \begin{equation}
 \dfrac{d\overline{\sigma}}{d\Omega_{f}}=\sum_{s=-\infty}^{+\infty}\dfrac{e^{4}}{8m_{\mu}Q_3 q^4} \frac{|\textbf{q}_{3}|^{2}}{|\textbf{q}_{1}|}\int \delta(p_4^2-m_{\mu}^2) d|\mathbf{q_3}| \sum_{s_{i}}  |\mathcal{M}_{fi}^s|^2.
 \end{equation}
The integration over $ d|\mathbf{q_3}| $ can be performed using the following familiar formula \cite{greiner}:
\begin{align}\label{familiarformula}
\int dxf(x)\delta(g(x))=\dfrac{f(x)}{|g'(x)|}\bigg|_{g(x)=0.}
\end{align}
Finally, we obtain for the laser-assisted DCS
\begin{equation}\label{dcswith}
\bigg(\frac{d\overline{\sigma}}{d\Omega_{f}}\bigg)^{e^{-}-\text{dressing}}=\sum_{s=-\infty}^{+\infty}\frac{d\overline{\sigma}^{(s)}}{d\Omega_{f}}=\sum_{s=-\infty}^{+\infty}\dfrac{e^{4}}{8m_{\mu}Q_3 q^4} \frac{|\textbf{q}_{3}|^{2}}{|\textbf{q}_{1}|}\frac{\sum_{s_{i}}|\mathcal{M}_{fi}^s|^{2}}{|g'(|\textbf{q}_{3}|)|},
\end{equation}
where
\begin{equation}
g'(|\mathbf{q_3}|)= 2(s\omega \cos(\theta_f)+ |\mathbf{q_1}| F(\theta_i,\theta_f,\phi_i,\phi_f)) - 2 \dfrac{|\mathbf{q_3}|}{Q_3}(Q_1 + m_{\mu} + s \omega).
  \end{equation}
The sum over spins will be converted into traces calculation as follows:
\begin{equation}\label{traces1}
\sum_{s_{i}}  |\mathcal{M}_{fi}^s|^2 =\text{Tr}[(\slashed{p}_3+m_{e}) \Gamma^\mu_s(\slashed{p}_1 +m_{e}) \overline{\Gamma}^\nu_s] \text{Tr}[(\slashed{p}_4+m_{\mu}) \gamma_\mu(\slashed{p}_2 +m_{\mu}) \gamma_\nu],
\end{equation}
where
\begin{equation}
\Gamma^\mu_s = \chi_0^\mu B_{s}(z)+\chi_1^\mu B_{1s}(z)+\chi_2^\mu B_{2s}(z),
\end{equation}
and
\begin{equation}
  \overline{\Gamma}^\nu_s  = \gamma^0 \Gamma_s^{\nu \dagger} \gamma^0 =  \overline{\chi}_0^\nu B_{s}^*(z)+ \overline{\chi}_1^\nu B_{1s}^*(z)+\overline{\chi}_2^\nu B_{2s}^*(z),
\end{equation}
with
\begin{equation}
\begin{split}
\overline{\chi}_0^\nu  &=  \gamma^{\nu}-e^{2}a^2 k^\nu\slashed{k}/[2(k.p_{1})(k.p_{3})],\\
\overline{\chi}_1^\nu  &=[e/2(k.p_1)]\slashed{a}_{1}\slashed{k}\gamma^{\nu}+[e/2(k.p_3)]\gamma^{\nu}\slashed{k}\slashed{a}_{1},\\
\overline{\chi}_2^\nu  &=[e/2(k.p_1)]\slashed{a}_{2}\slashed{k}\gamma^{\nu}+[e/2(k.p_3)]\gamma^{\nu}\slashed{k}\slashed{a}_{2}.
 \end{split}
 \end{equation}
The calculation of the traces appearing in eq.~(\ref{traces1}) is commonly performed with the help of FEYNCALC \cite{feyncalc1,feyncalc2,feyncalc3}. The result we obtained
is attached in the Appendix. In order to distinguish between them and to avoid any confusion, it would be very appropriate to consider the summed differential cross section (SDCS) in eq.~(\ref{dcswith}), $\Big(d\overline{\sigma}/d\Omega_{f}\Big)^{e^{-}-\text{dressing}}$, as the sum of discrete individual differential cross sections (IDCS), $d\overline{\sigma}^{(s)}/d\Omega_{f}$, for each photon exchange process.
\section{Muon-dressing effect}\label{dressing}
Up to now, we only consider the dressing of incident and scattered electrons. In this section and in order to highlight the effect of laser-dressing on the muon, we will take into account the relativistic dressing of both particles involved in the process (\ref{process}), the electron and the muon, and thus they will be described together by the Dirac-Volkov plane waves. This will introduce a new trace to be calculated and a new sum on the number of photons $n$ that will be exchanged between the muon and the laser field. The Dirac-Volkov wave function for the initial and final states of the dressed muon is such that
\begin{equation}\label{muwaves}
\begin{split}
\psi_{\mu^{-}}^{i,f}(y)&=\bigg[1+\dfrac{e\slashed{k}\slashed{A}'}{2(k.p_{2,4})}\bigg]\frac{u(p_{2,4},s_{2,4})}{\sqrt{2Q_{2,4}V}}\times e^{iS(q_{2,4},y)},
\end{split}
\end{equation}
where
\begin{equation}
\begin{split}
S(q_{2,4},y)=-(q_{2,4}.y)-\dfrac{e(a_1.p_{2,4})}{k.p_{2,4}}\sin(\phi')+\dfrac{e(a_2.p_{2,4})}{k.p_{2,4}}\cos(\phi').
\end{split}
\end{equation}
$A'(\phi')$ is the four-potential of the laser field felt by the muon
\begin{equation}\label{potlaser}
A'(\phi')=a^{\mu}_{1}\cos(\phi')+a^{\mu}_{2}\sin(\phi'),
\end{equation}
where $\phi'=(k.y)$ is the phase of the laser field. $Q_{2,4}$ is the total energy acquired by the muon in the presence of a laser field.\\
We do not need to repeat the steps leading to DCS because they have already been detailed in the previous section. Following the same procedure as before, we obtain for the unpolarized DCS
\begin{equation}\label{dressedmuondcs}
\bigg(\frac{d\overline{\sigma}}{d\Omega_{f}}\bigg)^{(e^{-},\mu^{-})-\text{dressing}}=\sum_{s,n=-\infty}^{+\infty}\frac{d\overline{\sigma}^{(s,n)}}{d\Omega_{f}}=\sum_{s,n=-\infty}^{+\infty}\dfrac{e^{4}}{8m_{\mu *}Q_3 q^4}\frac{|\textbf{q}_{3}|^2}{|\textbf{q}_{1}|}\frac{\sum_{s_i}|\mathcal{M}^{(s,n)}_{fi}|^{2}}{|h'(|\textbf{q}_{3}|)|},
\end{equation}
where $m_{\mu *}$ is the effective mass of the muon acquired inside the laser field. In this case, the energy-momentum conservation $q_{4}+q_{3}-q_{1}-q_{2}-(s+n)k=0$ must be satisfied. The term $\sum_{s_i}|\mathcal{M}^{(s,n)}_{fi}|^{2}$ is expressed by
\begin{equation}
\sum_{s_i}|\mathcal{M}^{(s,n)}_{fi}|^{2}=\text{Tr}[(\slashed{p}_3+m_{e}) \Gamma^\mu_s(\slashed{p}_1 +m_{e}) \overline{\Gamma}^\nu_s] \text{Tr}[(\slashed{p}_4+m_{\mu}) \Lambda_\mu^n(\slashed{p}_2 +m_{\mu}) \overline{\Lambda}_\nu^n],
\end{equation}
with
\begin{equation}
\Lambda_\mu^n=\chi '_{0\mu} B_{n}(z_{\mu^{-}})+\chi '_{1\mu} B_{1n}(z_{\mu^{-}})+\chi '_{2\mu} B_{2n}(z_{\mu^{-}}),
\end{equation}
and
\begin{equation}
\begin{split}
\chi '_{0\mu} &=\gamma_{\mu}-e^{2}a^2k_\mu \slashed{k}/[2(k.p_{2})(k.p_{4})],\\
\chi '_{1\mu} &=[e/2(k.p_2)] \gamma_{\mu}\slashed{k}\slashed{a}_{1}+[e/2(k.p_4)]\slashed{a}_{1}\slashed{k}\gamma_{\mu},\\
\chi '_{2\mu} &=[e/2(k.p_2)] \gamma_{\mu}\slashed{k}\slashed{a}_{2}+[e/2(k.p_4)]\slashed{a}_{2}\slashed{k}\gamma_{\mu}.
\end{split}
\end{equation}
$z_{\mu^{-}}$ is the argument of the new ordinary Bessel functions.
The quantity $h'(|\textbf{q}_{3}|)|)$ in eq.~(\ref{dressedmuondcs}) is given by
\begin{equation}
h'(|\textbf{q}_{3}|)|)=2\Big((s+n)\omega \cos(\theta_f)+ |\mathbf{q_1}| F(\theta_i,\theta_f,\phi_i,\phi_f)\Big) - 2 \dfrac{|\mathbf{q_3}|}{Q_3}\Big(Q_1 + m_{\mu *} + (s+n) \omega\Big).
\end{equation}
\section{Results and discussion}\label{sec:res}
We devote this section to the presentation and discussion of the numerical results obtained for the electron-muon relativistic scattering in the absence and presence of the laser field. The experimentally measurable quantity that is most important in scattering processes is the DCS, which expresses the probability of the event under consideration. Here, the DCS is derived with respect to the solid angle of the outgoing electron $\Omega_{f}$ and is evaluated in the muon laboratory frame where the muon is at rest with an initial energy $E_2=m_\mu$. For the geometry, we set both the initial and final electrons in a general geometry with spherical coordinates $\theta_i$, $\theta_f$, $\phi_i$ and $\phi_f$. These angles were chosen, in all presented results, such that $\theta_i=\phi_i$, $-180^{\circ}\leq\theta_f\leq 180^{\circ}$ and $\phi_f=\phi_i+90^{\circ}$. The latter geometry is chosen because it has been found to generate good and consistent results. The momentum of the final muon can be deduced from the other ones by using the momentum conservation relationship.
Taking into account relativity and spin effects, we choose the kinetic energy of the incoming electron as (unless otherwise stated) $E^{\text{kin}}_e=10^6~\text{eV}$.
\subsection{Without laser field}
\begin{figure}[h!]
 \centering
\includegraphics[scale=0.504]{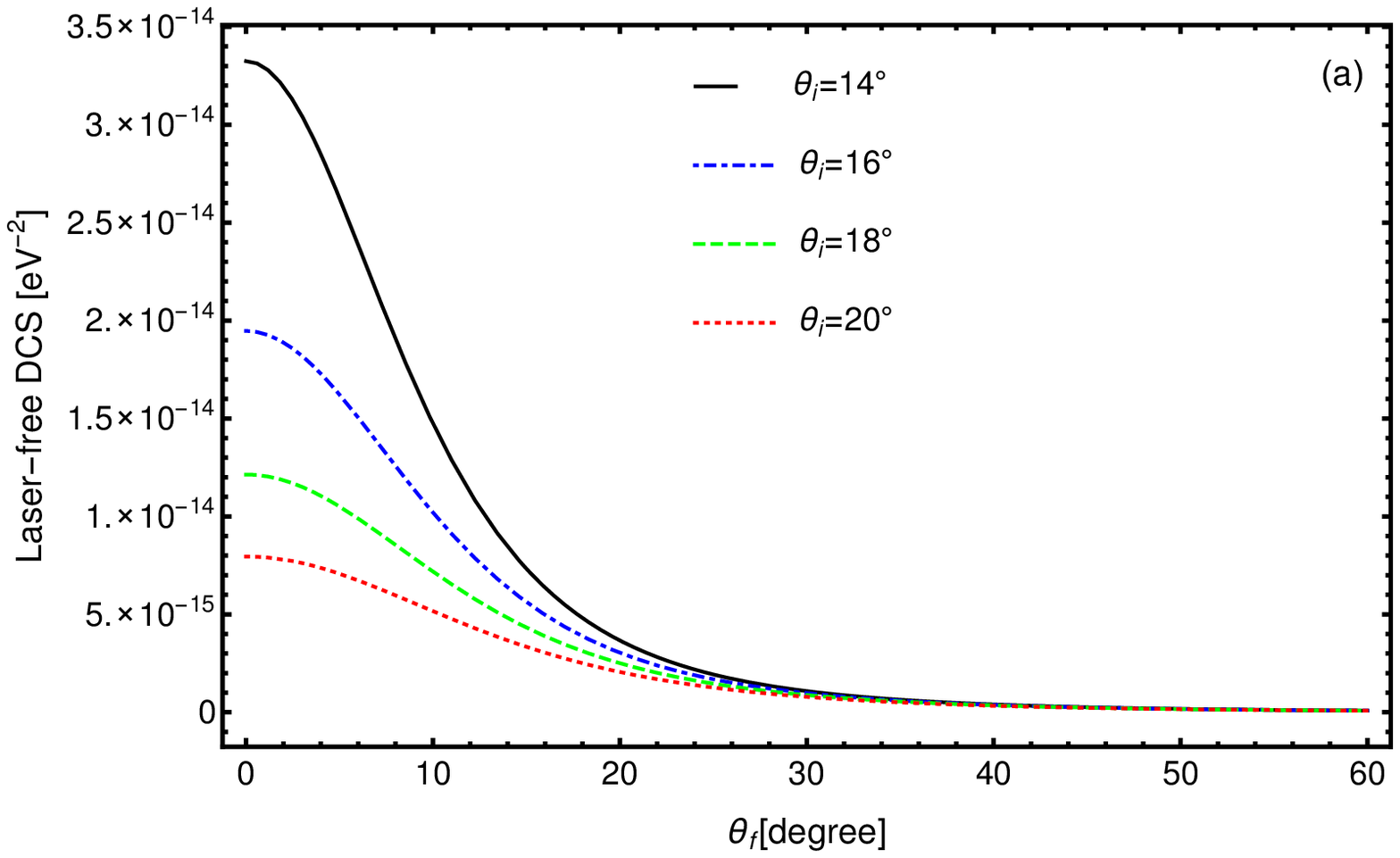}\hspace*{0.11cm}
\includegraphics[scale=0.504]{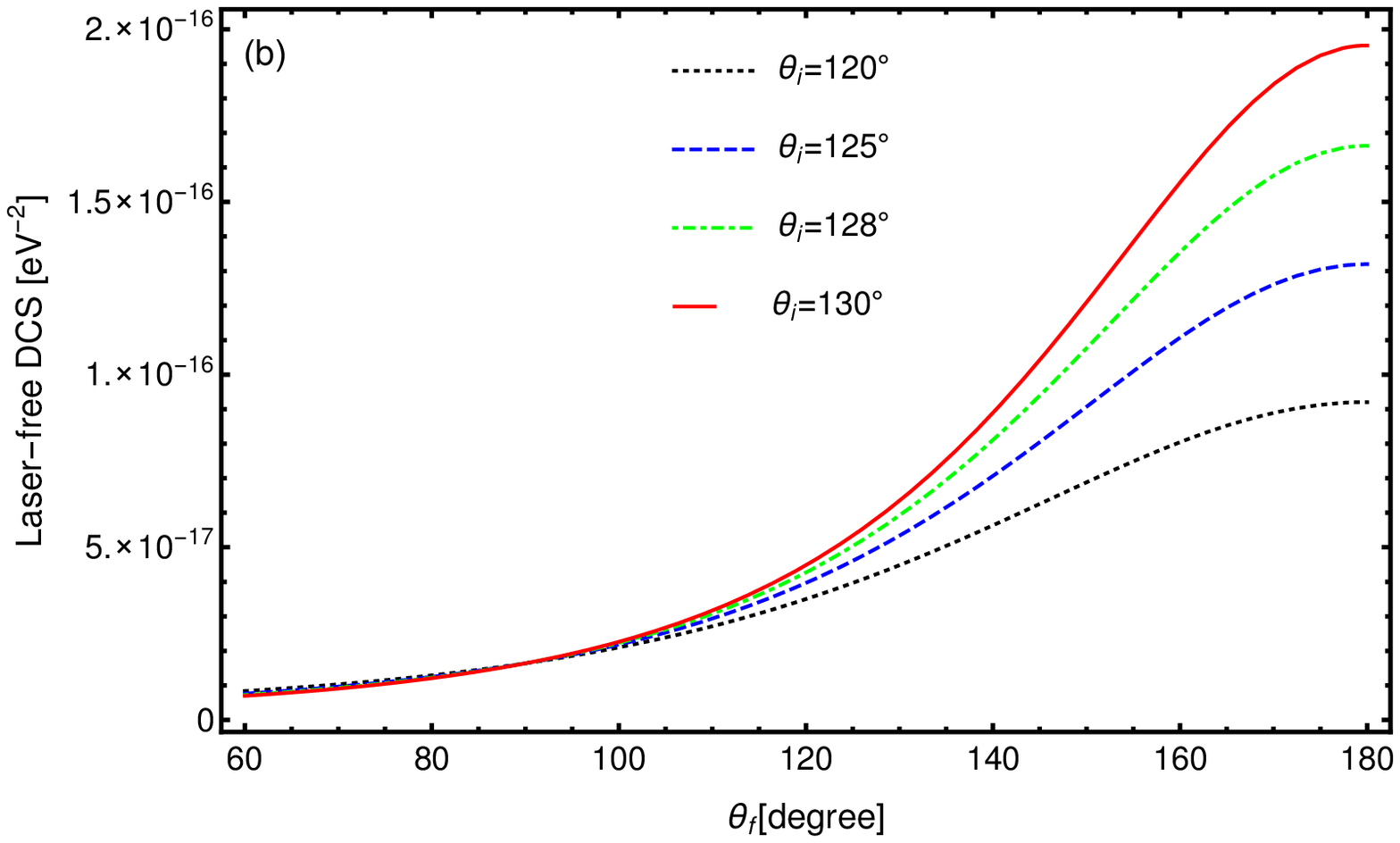}
\caption{The variations of the laser-free DCS as a function of the scattering angle $\theta_{f}$, for (a) small incident angles $\theta_{i}$ and (b) large incident angles $\theta_{i}$.}\label{fig0102}
\end{figure}

\begin{figure}[h!]
 \centering
\includegraphics[scale=0.5]{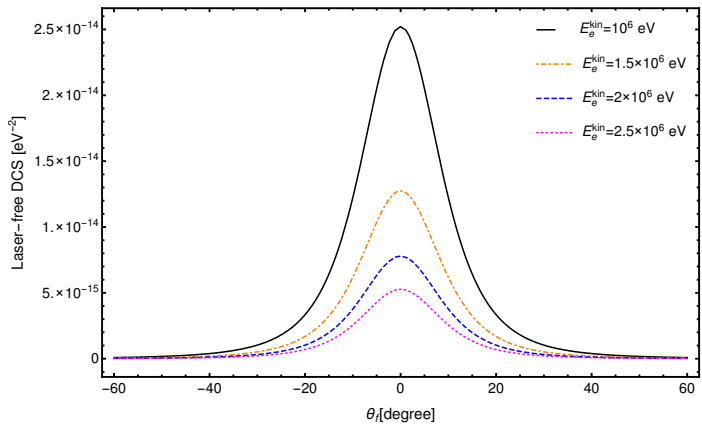}
\caption{The variations of the laser-free DCS as a function of the scattering angle $\theta_{f}$ for different kinetic energies of the incident electron. The incident angle is $\theta_{i}=15^{\circ}$.}\label{fig03}
\end{figure}

The parameters on which the laser-free DCS of the electron-muon scattering is dependent are the total energy of the incoming electron and various spherical coordinates. We control the initial parameters such as the kinetic energy of the incoming electron and its incident angle $\theta_i$, and we can see the effect of each on the DCS. As for the final parameters, we have no idea and we cannot control them. For the angle $\phi_f$, we found that the DCS does not change with it and gives a constant value. To obtain information about the final scattering angle $\theta_f$ at which the electron will be outgoing, it is necessary to study the variations of the DCS in terms of $\theta_f$ as shown in figure~\ref{fig0102} for different initial angles $\theta_i$.
For small initial angles, we notice that as the initial angle decreases $(\theta_i \rightarrow 0^{\circ})$, the DCS increases with a maximum value always at angle $\theta_f=0^{\circ}$. For large initial angles, we see that as the initial angle increases $(\theta_i \rightarrow 180^{\circ})$, the DCS increases with a maximum value at $\theta_f=180^{\circ}$. Whenever the initial angle tends to $0^{\circ}$ (forward scattering) or $180^{\circ}$ (backward scattering), the DCS is very considerable. If the electron comes in at a small initial angle $0^{\circ}\leq\theta_i<90^{\circ}$, there is a high probability that it comes out at an angle $\theta_f=0^{\circ}$. But, if it comes in at a large initial angle $90^{\circ}<\theta_i\leq 180^{\circ}$, it is very likely to come out at an angle $\theta_f=180^{\circ}$. For $\theta_i=90^{\circ}$, we found that all scattering angles have the same probability.\\
Let us now see the effect of the incoming electron kinetic energy on the laser-free DCS. In figure~\ref{fig03}, we plot the changes of the laser-free DCS in terms of the final scattering angle $\theta_f$ for different kinetic energies of the incident electron $E^{\text{kin}}_e$.
It is very clear that the DCS is inversely proportional to the kinetic energy, which means that it decreases with increasing the kinetic energy of the incoming electron. This is a normal and expected behavior due to the unitarity of the S-matrix element. The relativity and spin effects have significantly contributed to the discrepancies that arise between relativistic and non-relativistic regimes. It is also noticeable that the maximum value of the DCS, for $\theta_i=15^{\circ}$, is steadily located around the angle $\theta_{f}=0^{\circ}$ whatever the kinetic energy of the incident electron.
\begin{figure}[h!]
\centering
\includegraphics[scale=0.5]{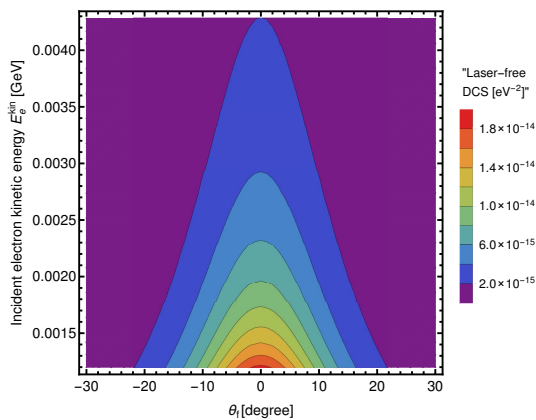}
\caption{The variations of the laser-free DCS as a function of both incident electron kinetic energy $E^{\text{kin}}_e$ and scattering angle $\theta_{f}$. The incident angle is $\theta_{i}=15^{\circ}$.}\label{fig04}
\end{figure}\\
To give a better and insightful overview of the laser-free DCS dependence on the incident electron kinetic energy and the final scattering angle $\theta_f$, we present in figure~\ref{fig04} a three-dimensional drawing (contour-plot) in which we represent the changes of the DCS in terms of two variables simultaneously, the kinetic energy of the incoming electron and the scattering angle $\theta_f$. From this figure, one can see that the curve is sharply peaked around the angle $\theta_f=0^{\circ}$ and that the order of magnitude of the laser-free DCS decreases with increasing the kinetic energy of the incoming electron.
\subsection{With laser field}
In this subsection, we will present the results obtained in the case of laser-dressing only the electron without the muon in the electron-muon scattering process. Each charged particle changes its properties and can acquire new ones when it interacts with an external field. The external field considered here is the circularly polarized electromagnetic field, which can be provided in the laboratory with a laser device \cite{lasercp}. In our case, the laser-dressed electron acquires a new effective mass, momentum and energy in the presence of the laser field. Hence, when we insert the electron-muon scattering process in an external electromagnetic field, its cross section will of course be modified and affected. Now, the laser parameters are added to those on which the DCS depends. Exactly, we are talking about the laser field strength and frequency as well as the number of photons exchanged. If these three parameters are zero within the limit of vanishing field, the laser-assisted DCS reduces down to the laser-free DCS. This comparison, which can be done numerically, helps to verify the consistency and accuracy of the theoretical calculation. For the laser geometry, we note here that we have chosen the direction of the field wave vector $\textbf{k}$ to be along the $z$-axis, whereas the polarization vectors $\textbf{a}_1$ and $\textbf{a}_2$ perpendicular to $\textbf{k}$ are along the $x$- and $y$-axes, respectively.
\begin{figure}[h!]
 \centering
\includegraphics[scale=0.51]{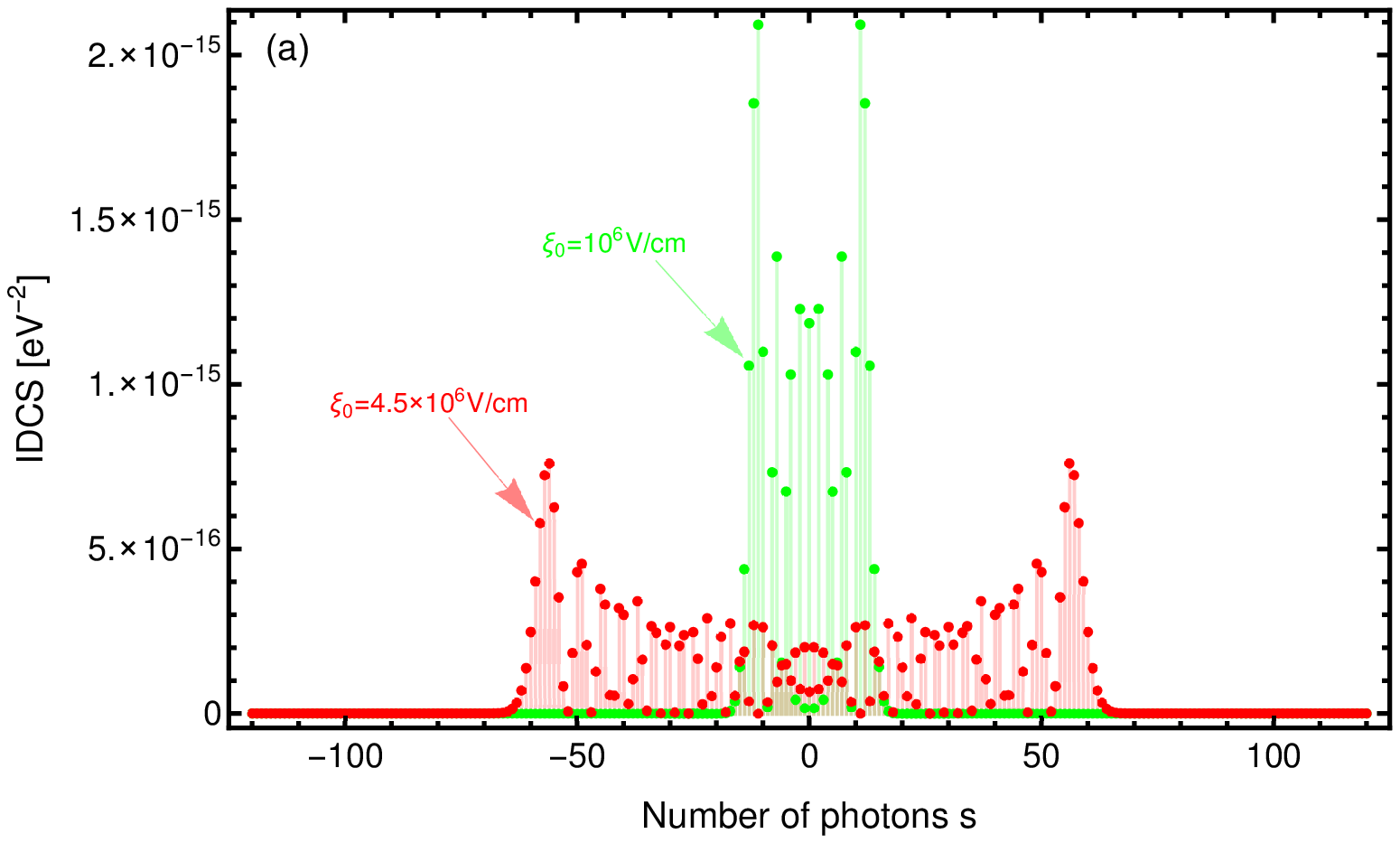}\hspace*{0.2cm}
\includegraphics[scale=0.5]{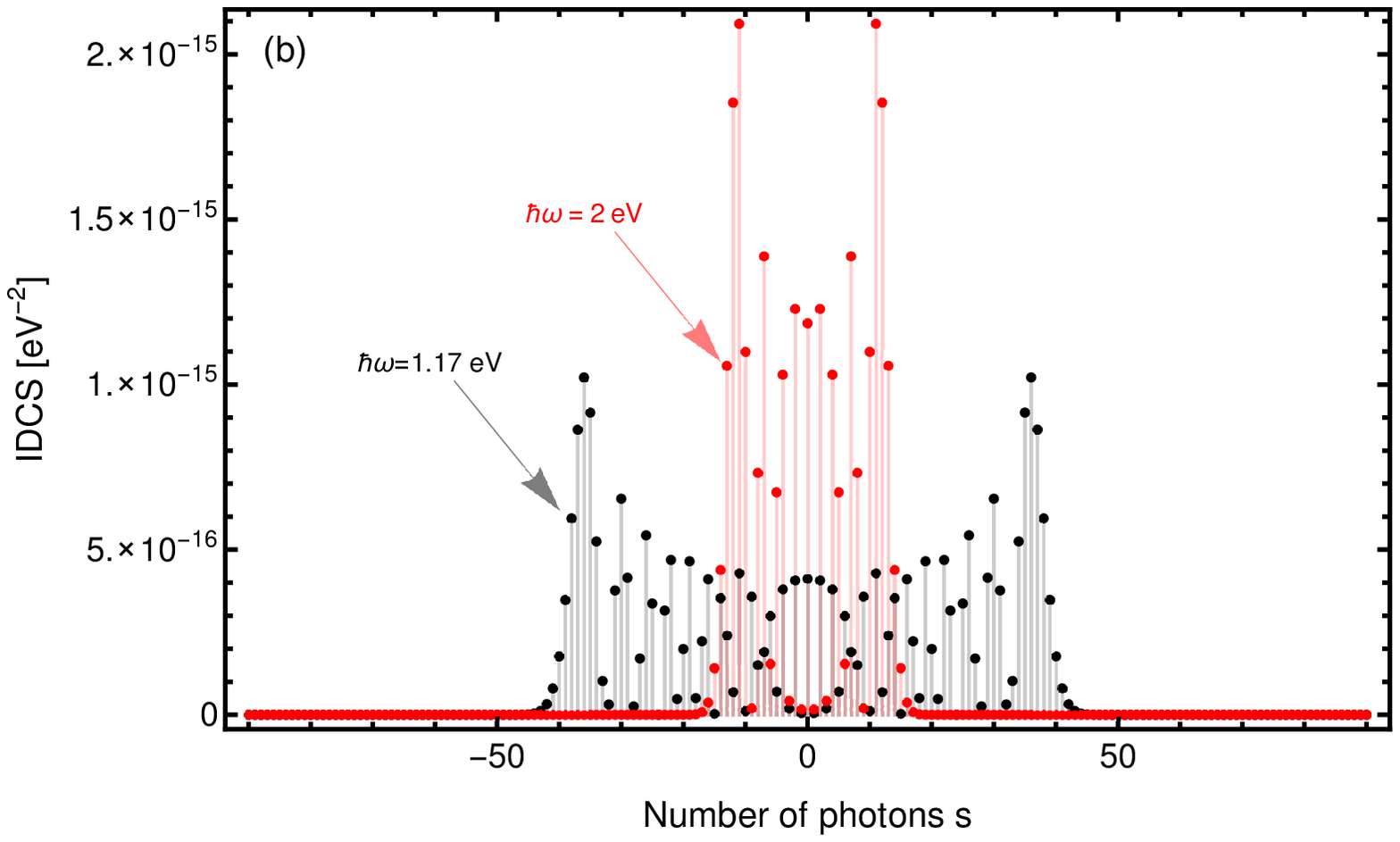}\par\vspace*{0.2cm}
\includegraphics[scale=0.5]{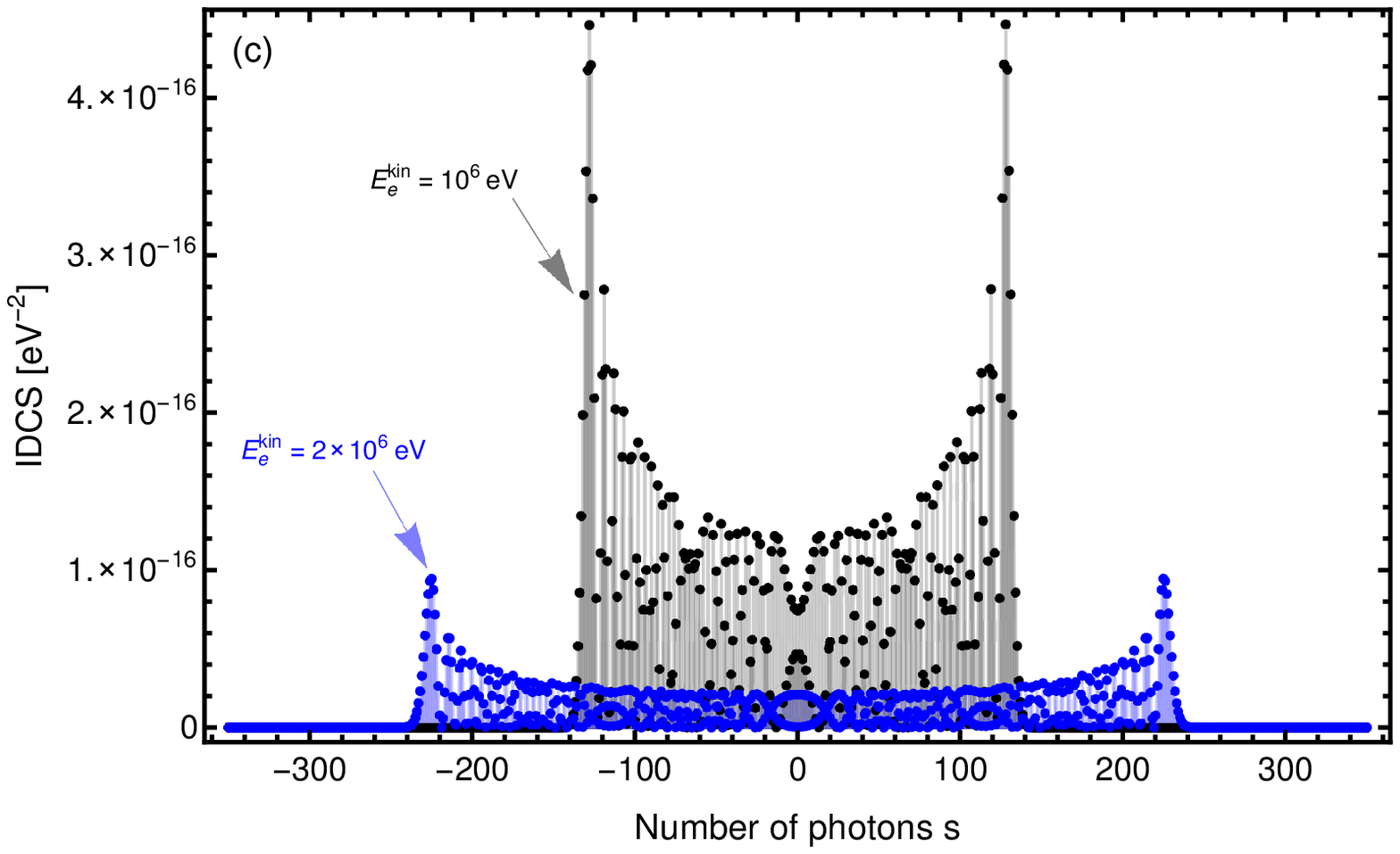}\hspace*{0.2cm}
\includegraphics[scale=0.5]{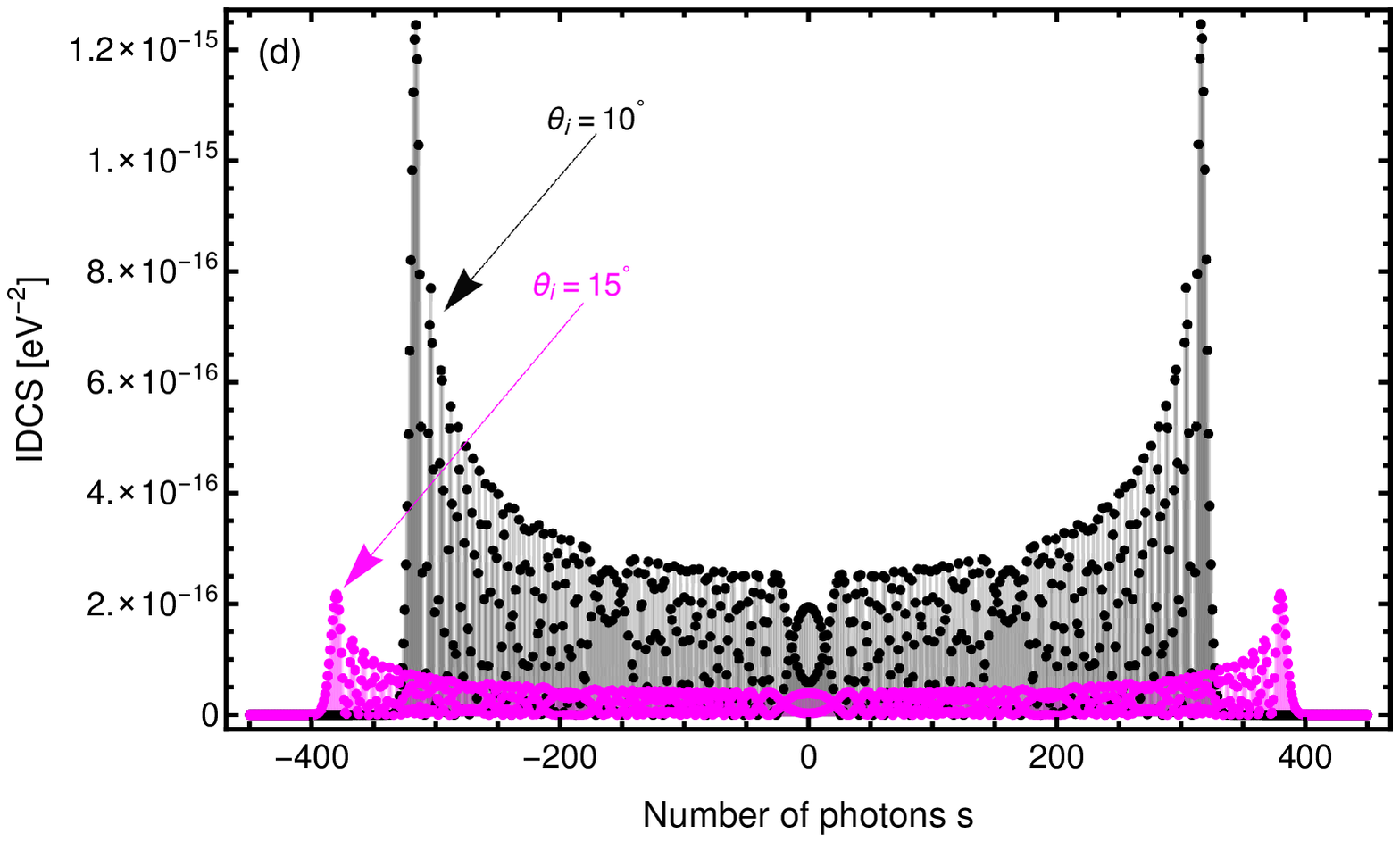}
\caption{The behavior of the IDCS, $d\overline{\sigma}^{(s)}/d\Omega_{f}$, versus the number of photons $s$. The different parameters are (a) $\hbar\omega=2$ eV, $\theta_{i}=15^{\circ}$ and $\theta_{f}=0^{\circ}$, (b) $\mathcal{E}_{0}=10^{6}$ V cm$^{-1}$, $\theta_{i}=15^{\circ}$ and $\theta_{f}=0^{\circ}$, (c) $\mathcal{E}_{0}=10^{7}$ V cm$^{-1}$, $\hbar\omega=2$ eV, $\theta_{i}=15^{\circ}$ and $\theta_{f}=0^{\circ}$ and (d) $\mathcal{E}_{0}=10^{7}$ V cm$^{-1}$, $\hbar\omega=1.17$ eV and $\theta_{f}=0^{\circ}$.}\label{envelopes}
\end{figure}
The important thing that indicates the interaction or non-interaction of the electron with the electromagnetic field is the process of photons exchange by emission and absorption. The exchange of a larger number of photons between the laser and the electron implies that the electron is interacting strongly with the laser field and vice versa. To examine the photon exchange process, we plot the variations the IDCS, $d\overline{\sigma}^{(s)}/d\Omega_{f}$, in terms of the number of photons exchanged $s$. As a result, we obtain envelopes as shown in figure~\ref{envelopes}. We note that all these envelopes are cut off at two edges which are symmetric with respect to $s=0$ axis. Figure~\ref{envelopes}(a) investigates the effect of laser field strength on the photon exchange process. It seems to us that the number of exchanged photons enhanced when we increased the field strength $\mathcal{E}_0$ from $10^6$ to $4.5\times 10^6$ V cm$^{-1}$. This indicates that the electron interacts significantly with the high-intensity laser field.
Concerning the effect of the laser frequency, we show in figure~\ref{envelopes}(b) the multiphoton process for two different frequencies. Through this figure, we can see that the electron exchanges few photons with the high-frequency laser ($\hbar\omega=2$ eV) compared to the low-frequency laser ($\hbar\omega=1.17$ eV). This means that the influence of laser diminishes at higher frequencies. Figure~\ref{envelopes}(c) illustrates the effect of the kinetic energy of the incident electron on its interaction with the laser field. From this figure, It appears that more photons were exchanged at higher kinetic energies than at lower energies. It means that the high kinetic energy of the electron contributed to enhance the interaction between the electron and the laser field. The effect of the initial angle $\theta_i$, at which the electron comes into the muon, was also studied; and the result is shown in figure~\ref{envelopes}(d). It is evident that there is a gap in the number of photons exchanged between the two initial angles, as more photons were exchanged in the case of wide initial angles. We have seen how all these initial parameters, whether those related to the laser, such as frequency and field strength, or those of the initial kinetic energy and geometry, play a major role in the way the electron interacts with the electromagnetic field.
\begin{figure}[hptb]
 \centering
\includegraphics[scale=0.53]{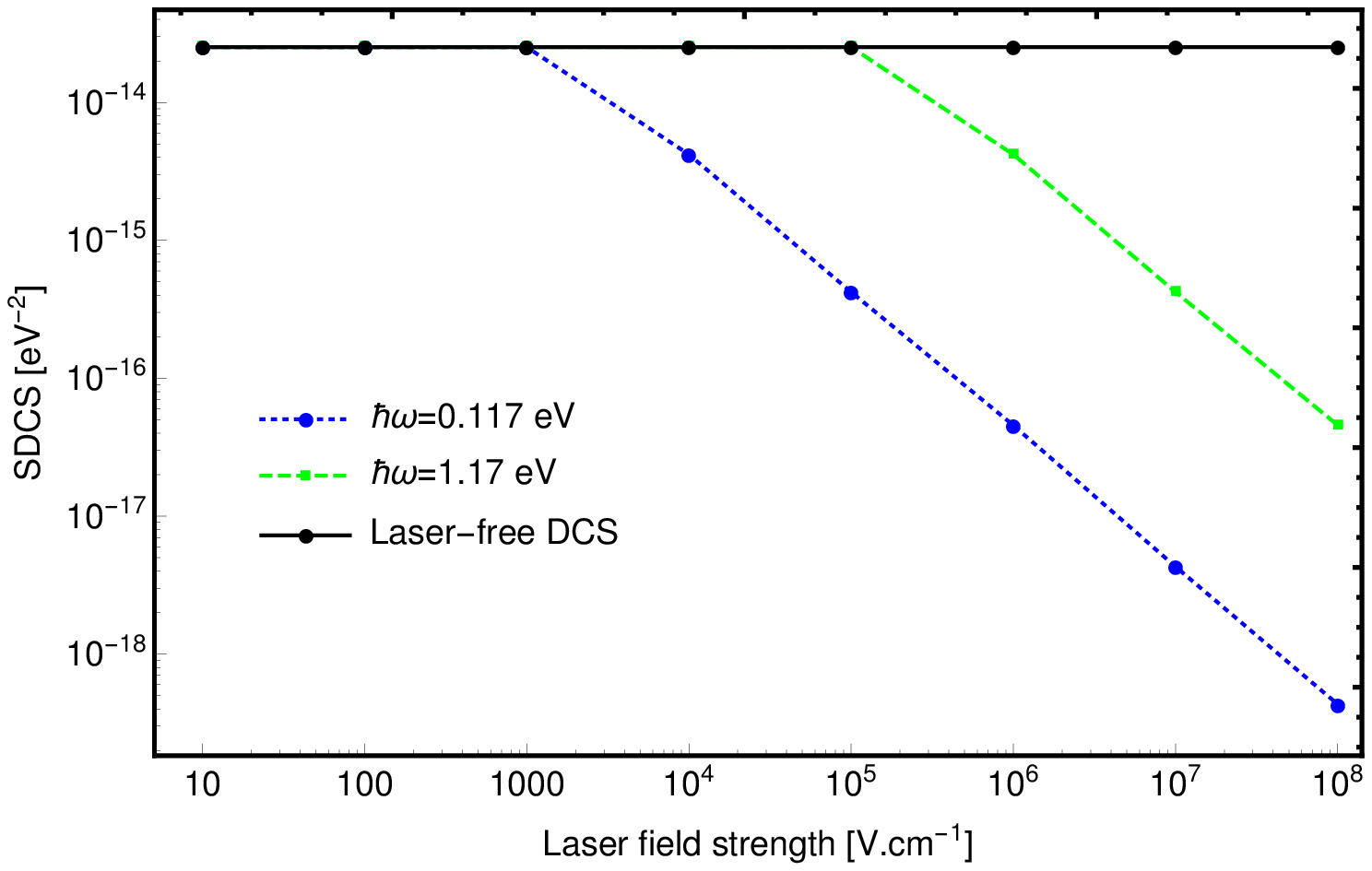}\hspace*{0.28cm}
\includegraphics[scale=0.55]{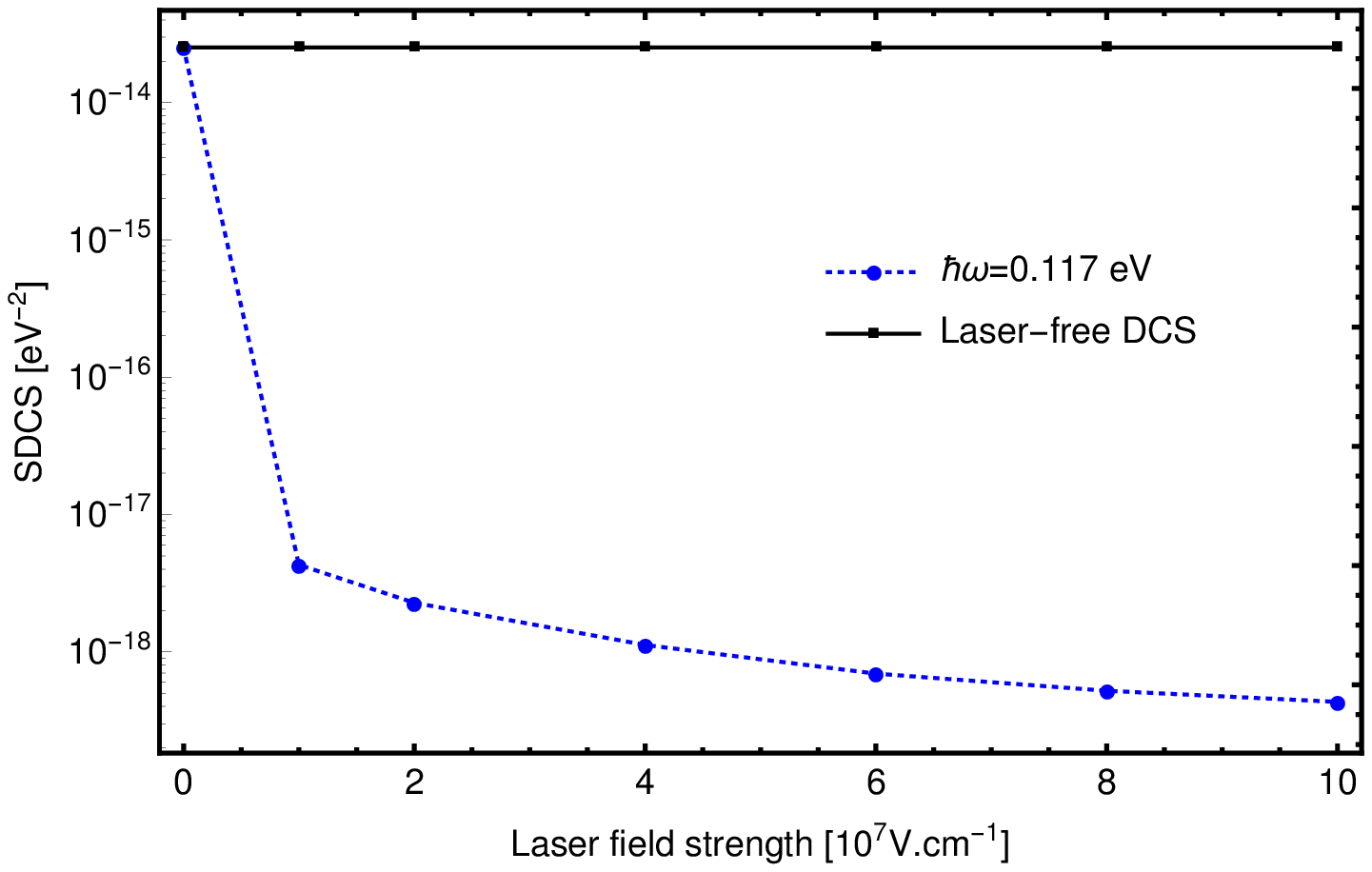}
\caption{The laser-assisted DCS (\ref{dcswith}) summed over $-10\leq s\leq 10$ as a function of the laser field strength for two laser frequencies. Panel (left) shows the range from $10$ to $10^8$ V cm$^{-1}$, while panel (right) shows the detail of the SDCS variations for the field strengths between $10^7$ and $10^8$ V cm$^{-1}$. The incident angle is $\theta_{i}=15^{\circ}$.}\label{fig0506}
\end{figure}\\
To further clarify the effect of laser parameters on the DCS of the muon-electron scattering, we show in figure~\ref{fig0506} the behavior of the SDCS (eq.~(\ref{dcswith})) in terms of laser field strength $\mathcal{E}_0$ for two different frequencies. From this figure, we can see that the laser, at its high intensities, contributed significantly to reducing the DCS. Also, the effect of the laser is dependent on the frequency used, as we see that the low-frequency laser affects the DCS faster than the high-frequency laser.
\subsection{Muon-dressing effect}
The purpose of this section is to present and discuss the results obtained in the case of the dressed muon, and to compare them with those obtained in the case where we dressed only the electron without the muon in the initial and final states. This is to determine at which field strengths the muon dressing has an effect on the DCS. Taking into account the interaction of both the electron and the muon with the electromagnetic field in the initial and final states makes the theoretical calculation somewhat difficult and heavy, requiring very fast and high-resolution computers to extract the numerical and graphical results. Due to this fact, we will limit ourselves here to include, in table~\ref{tab1}, some relevant values for the DCS in case we dress only the electron or electron and muon together.
\begin{table}[t]
\centering
\caption{Values of the two DCSs without (eq.~(\ref{dcswith})) and with (eq.~(\ref{dressedmuondcs})) the dressing effect of muon for different laser field strengths. The laser frequency is $\hbar\omega=1.17~\text{eV}$. The incident angle is $\theta_{i}=15^{\circ}$. The two numbers of photons $s$ and $n$ are summed from $-10$ to $+10$.}
\begin{tabular}{p{5cm}p{5cm}p{6cm}}
 \hline
Field strength $\mathcal{E}_{0}$ (V cm$^{-1}$)  & $\bigg(\frac{d\overline{\sigma}}{d\Omega_{f}}\bigg)^{e^{-}\text{-dressing}(\ref{dcswith})}$ (eV$^{-2}$)& $\bigg(\frac{d\overline{\sigma}}{d\Omega_{f}}\bigg)^{(e^{-},\mu^{-})\text{-dressing}(\ref{dressedmuondcs})}$ (eV$^{-2}$) \\
 \hline
$10^{5}$ & $2.52149\times 10^{-14}$ & $2.52149\times 10^{-14}$ \\
$10^{6}$ & $4.19919\times 10^{-15}$ & $4.19919\times 10^{-15}$ \\
$10^{7}$ & $4.22187\times 10^{-16}$  & $4.22187\times 10^{-16}$  \\
$10^{8}$ & $4.566\times 10^{-17}$ &$4.566\times 10^{-17}$ \\
$10^{9}$ &$4.30249\times 10^{-18}$ & $6.02908\times 10^{-19}$ \\
$10^{10}$ &$4.96829\times10^{-19} $& $6.38937\times10^{-21}$ \\
  \hline
\end{tabular}
\label{tab1}
\end{table}
From table~\ref{tab1}, we can easily see that the two DCSs are the same at the first four field strengths ($10^{5}$, $10^{6}$, $10^{7}$ and $10^{8}$ V cm$^{-1}$), which means that the laser-dressing of muon has no effect yet. As soon as we reach the field strength value of $10^{9}$ V cm$^{-1}$, we notice that the two DCSs start to be significantly different. At this field strength and above $(\mathcal{E}_0\geq 10^{9}~\text{V cm}^{-1})$, the muon-dressing effect begins to appear. In the range of field strengths below this value, it is enough to dress only the electron in order to avoid the complexity of the theoretical calculation and the cumbersome expressions generated by the laser-dressing of the muon. But, if higher field strengths are used, the muon must be dressed and described by Dirac-Volkov function in order to take into account its interaction with the intense electromagnetic field. We can attribute this to the difference between the masses of the muon and the electron. This would support and justify the choice made previously by some authors \cite{du2018,du2018p} to study the same scattering process without taking into account the laser-dressing of muon at the field strength of $5.18\times 10^7~\text{V cm}^{-1}$. Recently, Dahiri \textit{et al} \cite{dahiri} considered the effect of proton dressing in the laser assisted electron-proton scattering and found that it begins to appear at laser field strengths greater than or equal to $10^{10}~\text{V cm}^{-1}$ due to the heavy mass of the proton. From table~\ref{tab1}, it is also clear to us that the laser-dressing of muon has rendered the DCS lower than before.
\section{Conclusion}\label{sec:conclusion}
In order to conclude, we have presented in this work a theoretical study of the elastic relativistic electron-muon scattering in the presence of a circularly polarized monochromatic laser field under two phases. In the first one, we have dressed only the electron without the muon in the initial and final states, and in the second one we have also added the dressing of muon in order to highlight its effect on the DCS. In the obtained results, we have shown the effect of the laser field strength and frequency, as well as the kinetic energy of the incoming electron and its initial angle on the DCS and the multiphoton process. Our conclusions are as follows. The DCS is affected by the laser field, as it decreases with increasing field strengths due to the multiphoton absorption and emission processes. The scattering process can exchange a number of photons with the laser field depending on the field parameters and electron impact energy. We found that the photon emission processes are equal to the photon absorption ones. Regarding the laser frequency, it is shown that the effect of the laser decreases at high frequencies. For the impact energy, it is noted that as the incident energy increases, the DCS becomes smaller.
Yet another important point concerned the effect of the muon dressing, we have demonstrated that the laser has absolutely no effect on the DCS as long as the field strength is less than $10^{9}~\text{V cm}^{-1}$ and can therefore be safely ignored. Apart from all this, we recognize that in this theoretical work, as carried out in the framework of QED, we have considered only one Feynman diagram where the intermediate propagator is a photon $\gamma$ and therefore it remains valid at the lowest-order of perturbation theory. However, as it is well known, there are two other diagrams involving the Higgs and $Z$-bosons propagators that must be taken into consideration in order to ensure a complete and satisfactory study. This sounds very interesting and will be the subject of an upcoming investigation.
\section*{Appendix}
The numerical evaluation of the two traces shown in eq. (\ref{traces1}) gives the following result:
\begin{equation}
\begin{split}
\sum_{s_{i}}  |\mathcal{M}_{fi}^s|^2 =&\mathcal{C}_{1}|B_{s}|^{2}+\mathcal{C}_{2}|B_{1s}|^{2}+\mathcal{C}_{3}|B_{2s}|^{2}+\mathcal{C}_{4}B_{s}B^{*}_{1s}+\mathcal{C}_{5}B_{1s}B^{*}_{s}\\
&+\mathcal{C}_{6}B_{s}B^{*}_{2s}+\mathcal{C}_{7}B_{2s}B^{*}_{s}+\mathcal{C}_{8}B_{1s}B^{*}_{2s}+\mathcal{C}_{9}B_{2s}B^{*}_{1s},
\end{split}
\end{equation}
where the argument $z$ of the various ordinary Bessel functions has been dropped for convenience and the nine coefficients $\mathcal{C}_{1}$, $\mathcal{C}_{2}$, $\mathcal{C}_{3}$, $\mathcal{C}_{4}$, $\mathcal{C}_{5}$, $\mathcal{C}_{6}$, $\mathcal{C}_{7}$, $\mathcal{C}_{8}$ and $\mathcal{C}_{9}$ are explicitly expressed, in terms of different scalar products, by
\begin{equation}
\begin{split}
\mathcal{C}_{1}=&\dfrac{8 m_\mu}{(k.p_{1})^2 (k.p_{3})^2}\big[2 (k.p_{1}) (k.p_{3})(a^2 e^2 (k.p_{3})^2 E_{1}-a^2 e^2 (k.p_{1})^2 E_{3} +(k.p_{1}) (k.p_{3})(a^2 e^2(-E_{1} \\
&+ E_{3} + 2Q_{1} - 2 Q_{3})+ 2 (-m_\mu E_{1} E_{3} + E_{3} (p_{4}.p_{1})+E_{1} (p_{4}.p_{3})+m_e^2(m_\mu - Q_{1} + Q_{3}))))\\
&-2 a^2 e^2 (k.p_{1})(k.p_{3})(a^2 e^2 (-(k.p_{1}) + (k.p_{3}))+ (k.p_{3})(-2m^2+2 E_{1} E_{3} + (p_{4}.p_{1})) \\
&+(k.p_{1})(2 m_e^2 - 2 E_{1} E_{3} + (p_{4}.p_{3}))) \omega +a^4 e^4 (k.p_{1})(k.p_{3})m_\mu \omega^2 + a^6 e^6(-(k.p_{1}) + (k.p_{3}))\omega^3\\
&+2((k.p_{1})(k.p_{3})m_\mu+a^2 e^2(-(k.p_{1})+ (k.p_{3})) \omega)((k.p_{3})|\textbf{q}_{3}| \cos(\theta_{f})(a^2 e^2 \omega + 2 (k.p_{1})|\textbf{q}_{1}| \\
&\times\cos(\theta_{i}))+ (k.p_{1})(a^2 e^2 |\textbf{q}_{1}| \omega \cos(\theta_{i})+ 2 (k.p_{3}) |\textbf{q}_{3}| \sin(\theta_{f}) \sin(\theta_{i})(|\textbf{q}_{1}| \cos(\varphi_{f})\cos(\varphi_{i})\\
&+|\textbf{q}_{1}| \sin(\varphi_{f})\sin(\varphi_{i})))) \big],
\end{split}
\end{equation}
\begin{equation}
\begin{split}
\mathcal{C}_{2}=&\dfrac{8 e^2 m_\mu }{(k.p_{1})^2(k.p_{3})^2}\big[a^2(2(k.p_{1}) (k.p_{3})((k.p_{1})^2(m_\mu -E_{1}+E_{3})+(k.p_{3})^2(m_\mu -E_{1}+E_{3})\\
&- 2 (k.p_{1})(k.p_{3}) (m_\mu-E_{1}+E_{3}+Q_{1}-Q_{3}))+2 (k.p_{1})((k.p_{1})-(k.p_{3}))(k.p_{3})(2m^2\\
&-2E_{1} E_{3} - (p_{4}.p_{1})+(p_{4}.p_{3}))\omega +a^4 e^4 ((k.p_{1})-(k.p_{3}))\omega^3)+2 a^2((k.p_{1})-(k.p_{3}))\\
&\times(k.p_{3})|\textbf{q}_{3}| \omega \cos(\theta_{f})(a^2 e^2 \omega + 2 (k.p_{1})|\textbf{q}_{1}| \cos(\theta_{i})) +2 (k.p_{1})\omega(a^4 e^2 ((k.p_{1}) - (k.p_{3}))\\
&\times |\textbf{q}_{1}| \omega \cos(\theta_{i})+2 (k.p_{3})(|\textbf{a}|^2 (k.p_{1}) |\textbf{q}_{3}|^2 \cos(\varphi_{f})^2 \sin(\theta_{f})^2+(a^2((k.p_{1}) - (k.p_{3})) |\textbf{q}_{1}|  + |\textbf{a}|^2 \\
&\times(2 (k.p_{1}) |\textbf{q}_{1}|-(k.p_{3})|\textbf{q}_{1}|-(k.p_{1}) |\textbf{q}_{1}|)) |\textbf{q}_{3}| \cos(\varphi_{f}) \cos(\varphi_{i}) \sin(\theta_{f }) \sin(\theta_{i})+|\textbf{q}_{1}| \sin(\theta_{i})\\
&\times(-|\textbf{a}|^2 (k.p_{3}) |\textbf{q}_{1}| \cos(\varphi_{i})^2 \sin(\theta_{i})+a^2((k.p_{1})-(k.p_{3})) |\textbf{q}_{3}| \sin(\theta_{f}) \sin(\varphi_{f}) \sin(\varphi_{i}))))\big],
\end{split}
\end{equation}
\begin{equation}
\begin{split}
\mathcal{C}_{3}=&\dfrac{8 e^2 m_\mu}{(k.p_{1})^2(k.p_{3})^2}\big[a^2 (2 (k.p_{1}) (k.p_{3})((k.p_{1})^2 (m_\mu - E_{1} + E_{3}) + (k.p_{3})^2(m_\mu - E_{1} + E_{3})\\
&-2 (k.p_{1})(k.p_{3})(m_\mu -E_{1}+E_{3}+Q_{1}-Q_{3}))+2 (k.p_{1})((k.p_{1})-(k.p_{3}))(k.p_{3})(2 m_e^2\\
&-2E_{1} E_{3} - (p_{4}.p_{1}) + (p_{4}.p_{3})) \omega + a^4 e^4 ((k.p_{1}) - (k.p_{3})) \omega^3) + 2 a^2 ((k.p_{1}) - (k.p_{3})) (k.p_{3})\\
& \times|\textbf{q}_{3}| \omega \cos(\theta_{f}) (a^2 e^2 \omega +2 (k.p_{1}) |\textbf{q}_{1}| \cos(\theta_{i}))+ 2 (k.p_{1}) \omega (a^4 e^2 ((k.p_{1}) - (k.p_{3})) |\textbf{q}_{1}| \omega \cos(\theta_{i}) \\
&+2 (k.p_{3})(|\textbf{q}_{3}| \sin(\theta_{f}) (a^2 ((k.p_{1})-(k.p_{3})) |\textbf{q}_{1}| \cos(\varphi_{f}) \cos(\varphi_{i}) \sin(\theta_{i})+|\textbf{a}|^2 (k.p_{1})\\
&\times |\textbf{q}_{3}| \sin(\theta_{f})\sin(\varphi_{f})^2)+(a^2+|\textbf{a}|^2) ((k.p_{1})-(k.p_{3})) |\textbf{q}_{1}| |\textbf{q}_{3}| \sin(\theta_{f}) \sin(\theta_{i}) \sin(\varphi_{f}) \sin(\varphi_{i})\\
&-|\textbf{a}|^2 (k.p_{3}) |\textbf{q}_{1}|^2 \sin(\theta_{i})^2 \sin(\varphi_{i})^2))\big],
\end{split}
\end{equation}
\begin{equation}
\begin{split}
\mathcal{C}_{4}=&\mathcal{C}_{5}=\dfrac{8 |\textbf{a}| e m_\mu }{(k.p_{1}) (k.p_{3})}\big[|\textbf{q}_{3}| (2 (k.p_{1})((k.p_{1})(m_\mu - E_{1})- (k.p_{3})(m_\mu + E_{3}))+(a^2 e^2(3(k.p_{1})\\
&- (k.p_{3}))-2 (k.p_{1})(p_{4}.p_{1}))\omega)\cos(\varphi_{f})\sin(\theta_{f})-(2 ((k.p_{1})- (k.p_{3}))(k.p_{3})(m_\mu + E_{3})|\textbf{q}_{1}|\\
&-2 (k.p_{1})(k.p_{3})(E_{1} + E_{3}) |\textbf{q}_{1}| +(2(k.p_{3})(p_{4}.p_{3})|\textbf{q}_{1}|+a^2 e^2(-2 (k.p_{1})|\textbf{q}_{1}| + 2(k.p_{3}) |\textbf{q}_{1}|\\
&+ (k.p_{1})|\textbf{q}_{1}|+(k.p_{3})|\textbf{q}_{1}|)) \omega) \cos(\varphi_{i}) \sin(\theta_{i})\big],
\end{split}
\end{equation}
\begin{equation}
\begin{split}
\mathcal{C}_{6}=&\mathcal{C}_{7}=\dfrac{8~|\textbf{a}|~e m_\mu}{(k.p_{1}) (k.p_{3})}\big[|\textbf{q}_{3}| (2(k.p_{1})((k.p_{1})(m_\mu - E_{1})-(k.p_{3})(m_\mu + E_{3}))+(a^2 e^2 (3 (k.p_{1})\\
&-(k.p_{3}))-2 (k.p_{1})(p_{4}.p_{1}))\omega) \sin(\theta_{f}) \sin(\varphi_{f})+  |\textbf{q}_{1}| ((k.p_{1})(-2(k.p_{3})m_\mu + 2(k.p_{3}) E_{1}\\
&+a^2 e^2 \omega)+ (k.p_{3})(2 (k.p_{3})(m_\mu + E_{3})-(3 a^2 e^2 +2 (p_{4}.p_{3}))\omega)) \sin(\theta_{i})\sin(\varphi_{i})\big],
\end{split}
\end{equation}
\begin{equation}
\begin{split}
\mathcal{C}_{8}=&\mathcal{C}_{9}=\dfrac{16 |\textbf{a}|^2 e^2 m_\mu \omega }{(k.p_{1}) (k.p_{3})}\big[\cos(\varphi_{i}) \sin(\theta_{i}) ((2 (k.p_{1})|\textbf{q}_{1}| - (k.p_{3}) |\textbf{q}_{1}|- (k.p_{1}) |\textbf{q}_{1}|) |\textbf{q}_{3}| \sin(\theta_{f})\\
&\times \sin(\varphi_{f})-2(k.p_{3}) |\textbf{q}_{1}|^3  \sin(\theta_{i}) \sin(\varphi_{i})) +
 |\textbf{q}_{3}| \sin(\theta_{f}) ((k.p_{1}) |\textbf{q}_{3}| \sin(\theta_{f}) \sin(2\varphi_{f})\\
 & + ((k.p_{1})- (k.p_{3})) |\textbf{q}_{1}| \cos(\varphi_{f}) \sin(\theta_{i}) \sin(\varphi_{i}))\big].
\end{split}
\end{equation}

\end{document}